\documentclass[aps,pra,twocolumn,showpacs]{revtex4}
\usepackage[pdftex]{graphicx}
\usepackage{amssymb}
\usepackage{amsmath}
\usepackage{bm}
\usepackage{fancyref}

\DeclareGraphicsExtensions{.pdf}

\begin{document}

\title{Rotating a Bose-Einstein condensate by shaking an anharmonic axisymmetric magnetic potential}
\author{Seji Kang, J. Choi, S. W. Seo, W. J. Kwon, and Y. Shin}
\email{yishin@snu.ac.kr}

\affiliation{Center for Subwavelength Optics and Department of Physics and Astronomy, Seoul National University, Seoul 151-747, Korea}

\date{\today}

\begin{abstract}
 
We present an experimental method for rotating a Bose-Einstein condensate trapped in an axisymmetric magnetic potential. This method is based on the anharmonicity of the trapping potential, which couples the center-of-mass motion of the condensate to its internal motion. By circularly shaking the trapping potential, we generate a circular center-of-mass motion of the condensate around the trap center. The circulating condensate undergoes rotating shape deformation and eventually relaxes into a rotating condensate with a vortex lattice. We discuss the vortex nucleation mechanism and in particular, the role of the thermal cloud in the relaxation process. Finally, we investigate the dependence of the vortex nucleation on the elliptical polarization of the trap shaking. The response of the condensate is asymmetric with respect to the sign of the shaking polarization, demonstrating the gauge field effect due to the spin texture of the condensate in the magnetic potential.  

\end{abstract}


\maketitle

\section{Introduction}
	
Being irrotational, a superfluid shows a unique response to rotation, which involves quantized vortices. Since its first realization, a Bose-Einstein condensate (BEC) under rotation has been of great interest for the study of superfluidity and various vortex dynamics have been investigated in different rotation regimes~\cite{Fetter_rmp}. Many experimental methods were demonstrated for rotating trapped BECs, which include phase imprinting techniques based on optical transitions~\cite{Matthews99} or spin rotation~\cite{Leanhardt02,Choi12}, and applying rotating elliptical potentials~\cite{Madison_prl, Abo_Shaeer_science,Madison01,Raman_prl,Hodby02}. Rotating BECs were also produced by evaporative cooling of rotating thermal gases~\cite{Haljan_prl}. It is anticipated that in a rapid rotation limit a rotating BEC undergoes quantum phase transitions to strongly correlated quantum Hall states~\cite{Cooper01}, and there are ceaseless experimental efforts to reach the fast rotation regime~\cite{Schweikhard04,Bretin04,Roncaglia11}. 

In this paper, we present a new experimental method for rotating a trapped BEC. A trapped BEC can have angular momentum in two forms: the center-of-mass (CM) motion around the trap center and the internal motion such as surface oscillations and quantized vortices. The CM motion of the condensate can be easily generated, e.g. by shaking the trapping potential, to inject a large amount of angular momentum. However, in a harmonic potential, which is typically used in experiments, the CM motion is completely decoupled from the internal motion of the condensate and it is impossible to transfer the angular momentum associated with the CM motion to the internal rotation of the condensate. The rotating method described in this work is based on the anharmonicity of a trapping potential which provides the coupling between the CM motion and the internal motion of a trapped condensate~\cite{Marzlin_pra,Ott_prl,Lundh_prl,Guan_pra}. We drive the condensate to circulate around the trap center by circularly shaking the trapping potential and observe that the circulating condensate relaxes into a rotating condensate with a vortex lattice. This is indeed analogous to what we typically do to rotate liquid in a glass without a spoon. This method is very simple, requiring no sophisticated technique such as dynamically deforming the trapping potential.

This paper is organized as follows. In Sec.~II, we introduce our anharmonic axisymmetric trapping potential which is formed by a quadrupole magnetic field and describe the experimental procedure for generating a circular CM motion of a trapped condensate. In Sec.~III, we present the temporal evolution of the circulating condensate and discuss its relaxation to a rotating condensate with many vortices. The centrifugal effect and the role of the thermal cloud are found to be important in the relaxation process. In Sec.~IV, the dependence of the vortex nucleation on the elliptical polarization of the trap shaking is investigated. The condensate shows an asymmetric response with respect to the sign of the polarization of the driving, which is explained as the effect of the gauge field arising from the spin texture of the condensate in the magnetic potential~\cite{Choi_prl}. Finally, a summary is presented in Sec.~V.

\section{Experiment}

\begin{figure}
\includegraphics[width=8.0cm]{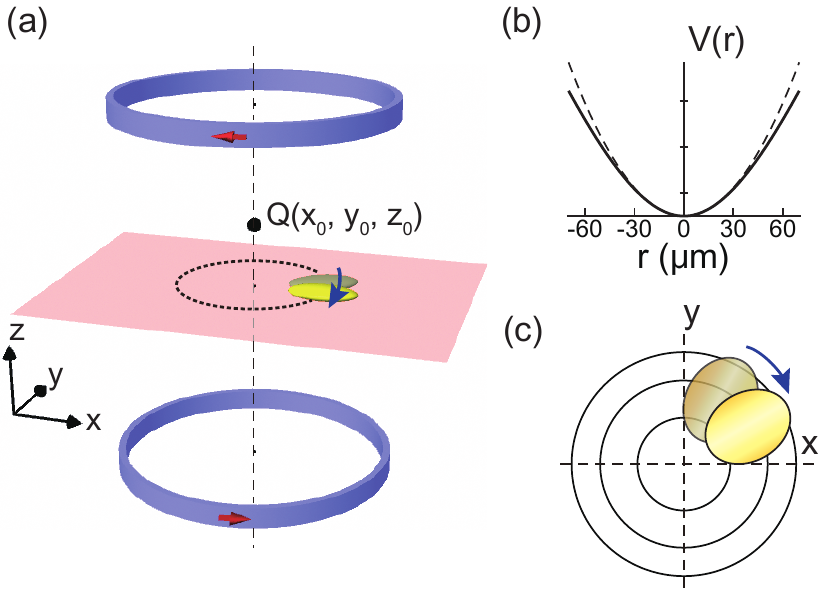}
    \caption{(Color online) Circulating Bose-Einstein condensate in an anharmonic magnetic potential. (a) Schematic of the experimental setup. A condensate (yellow ellipsoid) is confined in an optical sheet (red plane) and a quadrupole magnetic field is generated by a pair of coils (blue rings). The red arrows indicate the current directions and $Q$ is the zero-field point. (b) Anharmonic radial trapping potential from the quadrupole magnetic field. The dashed line shows a harmonic potential having the same curvature at the trap center. (c) At an off-center position, the local curvatures of the anharmonic trapping potential are different for the radial and azimuthal directions. When the condensate circulates around the trap center, it experiences a rotating anisotropic potential. The center-of-mass motion (CM) of the condensate is coupled to its internal motion.}
\label{fig1}
\end{figure}	

Our experiment uses Bose-Einstein condensates of sodium atom in the $|F=1, m_F=-1\rangle$ state~\cite{Heo11}. We prepare a condensate in a pancake-shaped optical dipole trap with trapping frequencies of $\omega_{x, y, z}=2\pi\times(3.3,4.1,400)$~Hz and apply to the condensate a three-dimensional quadrupole magnetic field, 
\begin{equation}
\mathbf B = \frac{B_q}{2}(x\hat x + y\hat y -2z\hat z)-B_x \hat x-B_y\hat y +B_z\hat z.
\end{equation}
The position of the zero-field point is controlled by the bias field $\{ B_x, B_y, B_z\}$ as $\mathbf{Q}=\{x_0,y_0,z_0\}=\{2B_x/B_q, 2B_y/B_q, B_z/B_q \}$. The overall external potential can be described as  
\begin{eqnarray}
V(x, y, z)&=&\frac{m}{2}(\omega_x^2 x^2+ \omega_y^2 y^2+\omega_z^2 z^2)\nonumber\\
&+&\frac{\mu_B B_q}{4} \sqrt{(x-x_0)^2+(y-y_0)^2+4  (z-z_0)^2}\nonumber\\
&+&mgz,
\end{eqnarray}
where $m$ is the atomic mass, $\mu_B$ is the Bohr magneton, and $g$ is the gravitational acceleration. The axial field gradient is set to be $B_q=7.6$~G/cm, almost compensating the gravitational sag in the optical trap, and the position of the zero-field point is placed at $\mathbf{Q}\approx \{0,0,32(2)\}~\mu$m [Fig.~1(a)]. The transverse trapping frequency at the center of the hybrid trap is $\omega_0=\sqrt{ \frac{\mu_B B_q}{8 m z_0}+\omega_{x,y}^2}$. For dipole oscillations with small amplitude of 16$~\mu$m, the trapping frequency was measured to be $\omega_0=2\pi \times 42.5$~Hz. The trap anisotropy due to the weak optical potential is estimated to be less than 0.2\%.

For the oblate condensate which is confined to the $z=0$ plane, we approximate the radial trapping potential as
\begin{equation}
V_r(r)=2m\omega_0^2 z_0 \sqrt{r^2+4z_0^2},
\end{equation}
where $r=\sqrt{x^2+y^2}$. This potential behaves linear for large $r\gg z_0$, representing a highly anharmonic, axisymmetric trap for the condensate  [Fig.~1(b)].  For a typical condensate of $\sim 2\times 10^6$ atoms, the Thomas-Fermi radius $R\approx 40~\mu$m which is comparable to $z_0$.

\begin{figure}
\includegraphics[width=8.0cm]{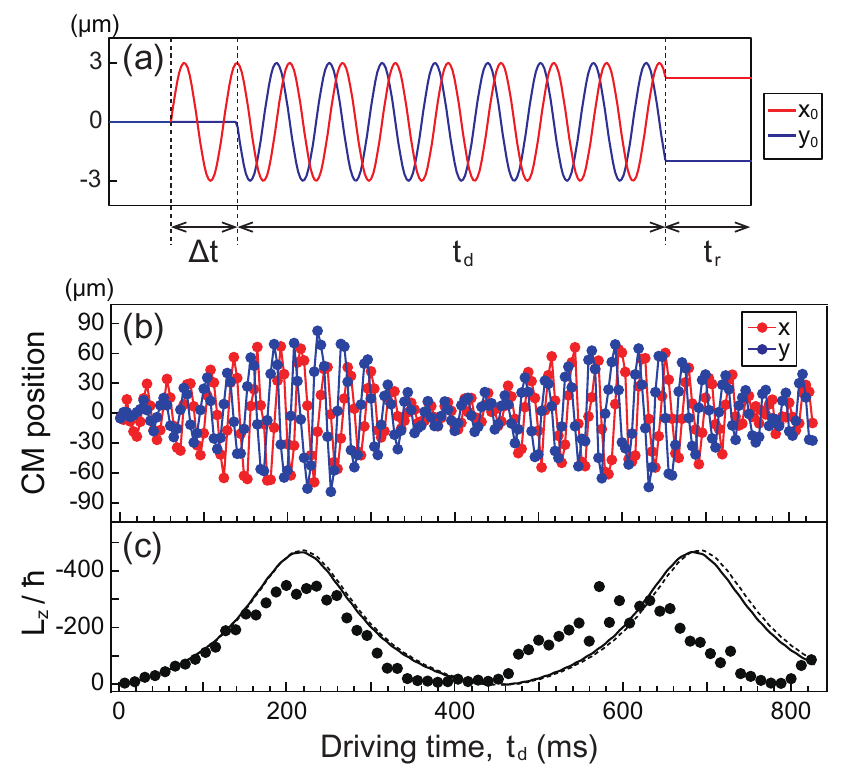}
    \caption{(Color online) Generation of a circular CM motion of the condensate. (a) The trapping potential is circularly shaken by applying a rotating bias field. The driving frequency is $\omega_m=2\pi\times39.5$~Hz close to the trapping frequency at the trap center $\omega_0=2\pi\times41.0$~Hz. (b) The temporal evolution of the CM position of the condensate  (single measurements for every 4~ms) and (c) the external angular momentum per particle, $L_z$, of the condensate. The CM velocity was determined from the position data in (b) and $L_z$ is displayed with 16 ms binning. The solid line denotes the numerical result from the single-particle simulation using Eq.~(4) and the dashed line shows the numerical result including the effective Lorentz force (see Sec.~IV). }
\label{fig1-2}
\end{figure}	

In order to generate a circular CM motion of the condensate, we apply a rotating bias field as $\{B_x(t),B_y(t)\}=\delta B \{\sin (\omega_m t),\cos (\omega_m t)\}$. The trap center correspondingly circulates around the origin as $\mathbf{d}(t)=\{x_0(t),y_0(t)\}=d \{\sin (\omega_m t),\cos (\omega_m t)\}$ with $d=2\delta B/B_q$. The turn-on sequence of the field modulations is displayed in terms of the trap position in Fig.~2(a). In the reference frame of the circulating magnetic trap, the equation of motion for an atom in the $z=0$ plane is given as
\begin{equation}
m\frac{\mathrm{d}^2\mathbf{r}'}{\mathrm{d}t^2}=-\frac{\mathrm{d}V_r(r')}{\mathrm{d}r'} +m\omega_m^2 \mathbf{d} (t)
\end{equation}
where $\mathbf{r}'=\{x-x_0,y-y_0\}$ and the second term is the inertial force due to the trap shaking. In the case of a harmonic trap with $V_r=\frac{1}{2}m\bar{\omega}^2r^2$, this rotating force will drive the atom to have a circular motion, whose radial position oscillates as $r'=[2\omega_m^2 d /(\bar{\omega}^2-\omega_m^2)] \sin (\frac{\bar{\omega}-\omega_m}{2} t)$. The angular momentum can be resonantly injected into the system.

Figure~2(b) shows the temporal evolution of the CM position of the condensate for $\omega_m=2\pi\times 39.5$~Hz and $d=3~\mu$m. An \textit{in situ} absorption image was taken every 4~ms of the driving and the CM position was determined from a two-dimensional Gaussian fitting.  The condensate develops a circular CM motion and the radius of the circular motion oscillates with a period of 400~ms. For $t_d=200$~ms driving, the radial position of the condensate increases up to about 65~$\mu$m and the external angular momentum per particle, associated with the CM motion, increases as large as $|L_z|\approx 350\hbar$, where $\hbar$ is the Planck constant divided by $2\pi$. We see that the measured CM trajectory is reasonably accounted for by the single-particle simulation using Eq.~(4)~[Fig.~2(c)]. The deviation for later time is understandable because the CM motion of the condensate having a finite spatial extent cannot be separable from its internal motion in the anharmonic potential~\cite{Ott_prl,Lundh_prl,Guan_pra}. In the following experiment, we employ the same driving procedure.

\section{Results}\label{Vortex}

\subsection{Surface mode excitations}

At an off-center position in the anharmonic trapping potential, the local curvature of the potential is not isotropic. For the external potential in Eq.~(3), the local trapping frequencies along the radial direction and the azimuthal direction are given as
\begin{eqnarray}
\omega_r&=&\sqrt{\frac{1}{m} \frac{\partial^2 V_r}{\partial r^2}}=\omega_0 \left[1+\frac{r^2}{4z_0^2}\right]^{-3/4} \nonumber\\
\omega_\theta&=&\sqrt{\frac{1}{m r}\frac{\partial V_r}{\partial r}}=\omega_0 \left[1+\frac{r^2}{4z_0^2}\right]^{-1/4},
\end{eqnarray}
respectively, and the local trap anisotropy $\epsilon=(\omega_\theta^2-\omega_r^2)/(\omega_\theta^2+\omega_r^2)\geq 0$. Therefore, the circulating condensate would experience a rotating anisotropic trapping potential in its CM reference frame~[Fig.~1(c)]. As the radial position of the condensate increases under the driving, the local trap anisotropy gradually increases. The rotating anisotropic potential would cause surface mode excitations~\cite{Stringari96,Onofrio_prl}, possibly leading to vortex nucleation~\cite{Madison_prl, Abo_Shaeer_science,Madison01,Raman_prl,Hodby02,Dalfovo_pra2,Recati01,Sinha01,Tsubota02_pra,Kasamatsu03pra,Lundh03,Lobo_prl,Parker05}.

\begin{figure}[b]
\includegraphics[width=8.3cm]{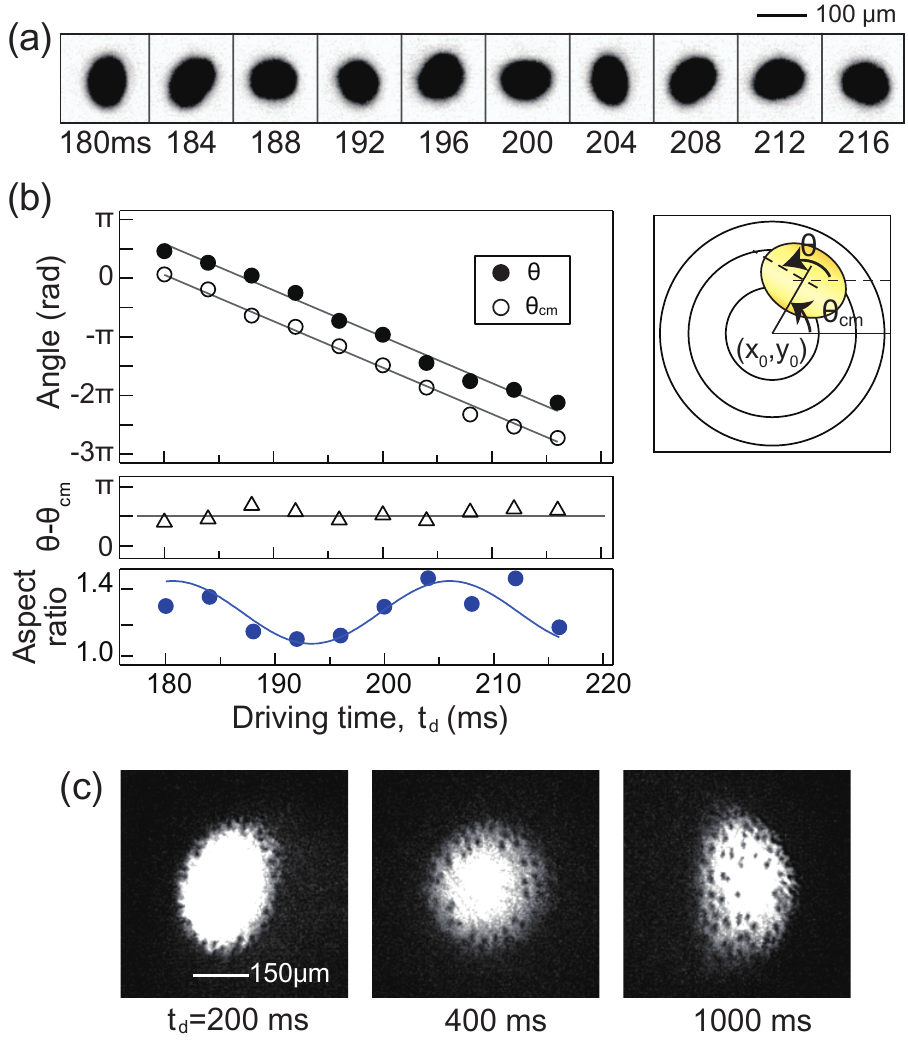}
\caption{(Color online) Surface mode excitations in the circulating condensate. (a) \textit{In situ} absorption images of the condensates for various driving times $t_d$. (b) Time evolution of the shape deformation of the condensate. $\theta$ and $\theta_{cm}$ denote the orientation of the long axis of the condensate and the azimuthal angle of the CM position, respectively. The aspect ratio shows small oscillations. The solid lines are guide lines for the eye. (c) Images after 10 ms time-of-flight expansion. Vortices are nucleated in the boundary region of the condensate.}
\end{figure}	

Figure 3(a) shows a series of \textit{in situ} absorption images of the condensate for $t_d=180\sim 216$~ms. In this period, the radial position of the condensate is $r\approx65~\mu$m, giving $\epsilon=0.34$ with $\omega_\theta/\omega_0=0.84$ and $\omega_r/\omega_0=0.59$. We see that the circulating condensate is elliptically deformed and the axis of the deformation rotates with the same frequency of the condensate circulation~[Fig.~3(b)]. The long axis of the condensate is found to be along the azimuthal direction, which is orthogonal to the weak confining direction of the local anisotropic trapping potential. This is consistent with the theoretical prediction for the case where the rotating frequency $\Omega$ of the elliptical potential is higher than the trapping frequencies~\cite{Recati01}, which is true in our experiment for $\Omega \approx \omega_m > \omega_{\theta,r}$. The steady state solution for a condensate in a rotating elliptical harmonic potential predicts the aspect ratio of the deformed condensate to be 1.16 for our experimental condition with $\Omega/\omega_\theta\approx 1.11$ and $\epsilon=0.3$~\cite{Recati01}. We observe that the aspect ratio of the circulating  condensate oscillates around 1.25 [Fig.~3(b)], which is in reasonable agreement with the prediction.

We identify vortex nucleation in the circulating condensate by taking an absorption image after releasing the trapping potential and subsequent expansion [Fig.~3(c)]. Vortices are generated in the condensate boundary and gradually diffuse into the inner region. The vortex population increases for longer driving time. In previous experiments~\cite{Madison_prl,Abo_Shaeer_science,Madison01,Raman_prl,Hodby02} which investigated condensates in rotating elliptical potentials, the vortex nucleation mechanism was mainly associated with the dynamic instability via the quadrupole surface mode  resonance~\cite{Recati01,Sinha01,Tsubota02_pra,Lundh03,Lobo_prl,Kasamatsu03pra,Parker05}. 
 In our experiment, the rotating frequency $\Omega$ is almost two times higher than the resonance frequency of the quadrupole surface mode, $\Omega_2\approx \sqrt{\omega_\theta^2+\omega_r^2}/2=2\pi\times 21.7$~Hz, and the dynamic instability is not likely to be involved in the vortex nucleation. Furthermore, the observed oscillations of the aspect ratio of the condensate indicate that the deformed condensate is dynamically stable. This suggests that the vortex nucleation in the circulating condensate is driven by the energetic instability in the rotating potential.

\subsection{Relaxation of the circulating condensate}

After stopping the trap shaking, we investigate the relaxation of a freely circulating condensate. Figure~4(a) shows the time-of-flight images of the condensate for various relaxation times $t_r$ after 200~ms driving. In the early phase of the evolution, the condensate keeps circulating, maintaining its elliptically deformed shape. It is known that a particle in a rotating elliptical harmonic potential is dynamically unstable for $\omega_r<\Omega<\omega_\theta$~\cite{Guery00,Recati01,Madison01}. Although $\Omega=\omega_\theta$ in our case for the free circulation, close to the unstable regime, the condensate remains stable without showing any turbulent behavior. Over 500~ms, more vortices are nucleated up to $N_v\approx 40$ and the radius of the CM motion decreases gradually to about 45~$\mu$m.

\begin{figure}
\includegraphics[width=8.5cm]{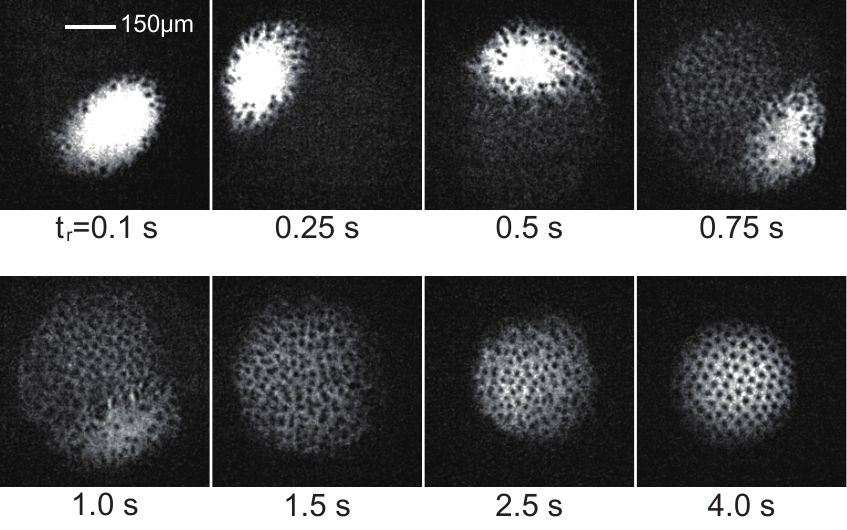}
    \caption{Relaxation of the circulating condensate in the anharmonic trapping potential. Time-of-flight images of the condensates for various relaxation times $t_r$ after 200~ms driving. The circulating condensate evolves into a rotating condensate with a triangular lattice of vortices.}
\end{figure}

As the relaxation proceeds, we observe that a faint cloud slowly forms over a large central area. The cloud grows in density and reveals a large number of vortices, which indicates that the newly formed cloud becomes a part of the condensate. This process is remarkably gentle and the elliptical region of the original circulating condensate is well preserved as a high-density part. The high-density part of the condensate gradually shrinks and becomes completely dissolved at $t_r\approx 1.5$~s. At this moment, we have a large round condensate containing many vortices whose spatial distribution is disordered. The vortex number is counted up to $N_v\approx200$. This state represents a rotating condensate with an irregular vortex lattice.

The development of the rotating condensate might be understood as a thermal equilibrating process in the reference frame corotating with the circulating condensate (Fig.~5). In the rotating reference frame, the effective trapping potential is given as $V^e(r)=V(r)-\frac{1}{2}m\omega_\theta^2 r^2$, including the centrifugal potential. Because of the anharmonicity of $V(r)$, $V^e(r)$ has a restricted trapping region with a global potential minimum at the center. The circulating condensate is depicted to be located at the edge of the trapping region with the help of the Coriolis force [Fig.~5(a)]. Since the circulating condensate has an effective chemical potential higher than the potential minimum, it would thermodynamically drive thermal atoms to condense at the trap center, which would result in a rotating condensate at the trap center. 

This is a kind of a thermal distillation process where a condensate in a metastable state is transferred to the true ground state via thermal atoms~\cite{Shin04}. Note that such a distillation process would not occur in a system using a harmonic potential, where the effective external potential for a circulating condensate is flat, having no potential minimum. When a quartic potential is employed, which provides stronger confinement at a larger radius, the effective potential has a ring geometry and a circulating condensate would relax into a giant vortex state~\cite{Lundh_pra,Kavoulakis_NJP}.

\begin{figure}[b]
\includegraphics[width=7cm]{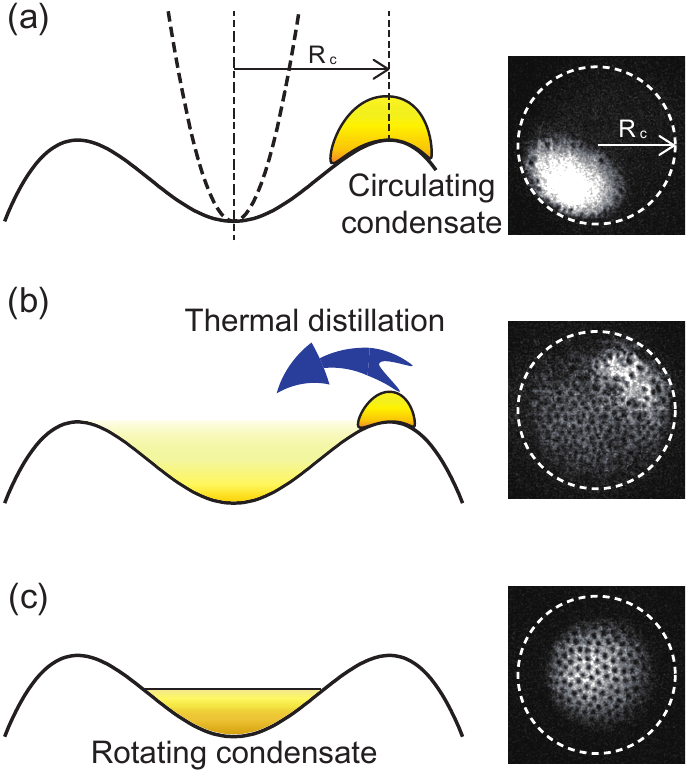}
\caption{(Color online) Thermal distillation process. (a) In a reference frame corotating with the circulating condensate, the condensate is located at the edge of the trapping region of the effective trapping potential (solid line). The dashed line shows the trapping potential in the lab frame. (b) The condensate thermodynamically drive thermal atoms to condense in the center region having the trapping potential minimum. (c) A rotating condensate is formed at the trap center. The right column displays the images corresponding to each step of the relaxation process. The dashed circles indicate the trap boundary determined by the centrifugal effect.}
\end{figure} 

In our experiment, the collision time of thermal atoms is estimated to be $\tau \sim 25$~ms for the sample temperature $T\sim 300$~nK~\cite{footnote3}. The growth of the rotating condensate happens on a time scale of $\sim20\tau$, which is comparable to the results in previous studies on Bose condensation~\cite{Miesner98}. The formation of the rotating condensate over a large area implies that the atomic influx from the circulating condensate is higher than the internal equilibration rate of the rotating condensate. 

The round boundary of the whole condensate, which is determined by the radial position of the circulating condensate, seems to reveal the trapping radius of the effective potential in the rotating reference frame [Fig.~5(a)]. From $\mathrm{d}V^e_r/\mathrm{d}r=0$, the trapping radius is given as $R_c=2z_0 \sqrt{(\omega_0^4/\omega_\theta^4)-1}$. When the spatial extent of a rotating condensate is limited by the centrifugal effect, the vortex number of the condensate would be
\begin{equation}
N_v^c\approx \frac{m}{\hbar}\omega R_c^2 =
\frac{4m z_0^2}{\hbar}\frac{\omega_0^4-\omega^4}{\omega^3},
\end{equation}
which is estimated from the circulation relation $\oint_{r=R_c} \mathrm{v}_s\cdot d\mathrm{r} =(h/m) N_v^c$, where $\mathrm{v}_s$ is the superfluid velocity. For $t_r=1$~s, the \textit{in situ} radius of the rotating condensate was measured to be about $50~\mu$m, suggesting $N_v^c\approx 200$. This number is comparable to the maximum vortex number observed in the experiment. The rotating frequency of the newly formed rotating condensate is forced to be set to the circulation frequency of the initial condensate.

\begin{figure}
\includegraphics[width=7cm]{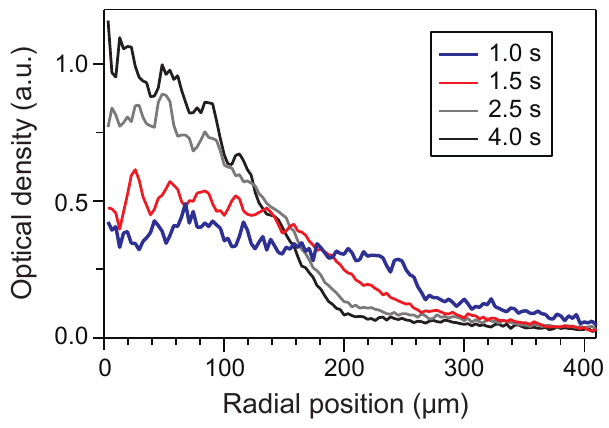}
    \caption{(Color online) Radial density distributions of the rotating condensates, which are obtained by azimuthally averaging the images in Fig.~4. For $t_r=1$~s, the high-density part, corresponding to the original circulating condensate, is excluded in the averaging. The density modulations in the center region result from density-depleted vortex cores.}
\end{figure} 

After the high-density part in the condensate vanishes, the rotating condensate with an irregular vortex lattice undergoes an equilibrating process. Crystalization into a triangular vortex lattice takes about 2~s, where the radial extent of the condensate decreases by about 25\% and the atomic density at the center increases by a factor of 2~(Fig.~6).  At $t_r=4$~s, the vortex number is $N_v\approx120$. One may expect an increase of the rotating frequency due to the reduction of the moment of inertia of the condensate, but the vortex areal density, i.e. the rotation frequency of the condensate shows no significant change during the evolution. The thermal fraction of the sample is as high as 50\% at this phase and thus, the interactions between the condensate and the thermal cloud must be important. The rotating thermal cloud might lock the condensate rotation in the equilibration process. With an initial state having a triangular lattice with about 100 vortices, the half lifetime of the vortices was measured to be 9~s.

\subsection{Rotating thermal cloud}

Next, we present the relaxation of the condensate after 400~ms driving. This case is special in that the CM motion of the condensate almost stops after one oscillation of the radial position [Fig.~2(c)]. Right after the driving, the condensate shows a round shape and contains about 30 vortices in its boundary region. As the hold time increases, we observe that more vortices come in and migrate to the inner region of the condensate, and finally, a triangular lattice of about 60 vortices forms after 2~s [Fig.~6(a)]. Since the angular momentum of the condensate is increased during the evolution~[Fig.~6(b)]~\cite{footnote}, it is obvious that the thermal cloud surrounding the condensate rotates faster than the condensate at the beginning. During the driving period, the thermal cloud is not only subject to the applied rotating force, but also dragged by the circulating condensate. We note that the time scale of the vortex lattice crystalization is similar to that in the previous case with $t_d=200$~ms. 

\begin{figure}
\includegraphics[width=8.3cm]{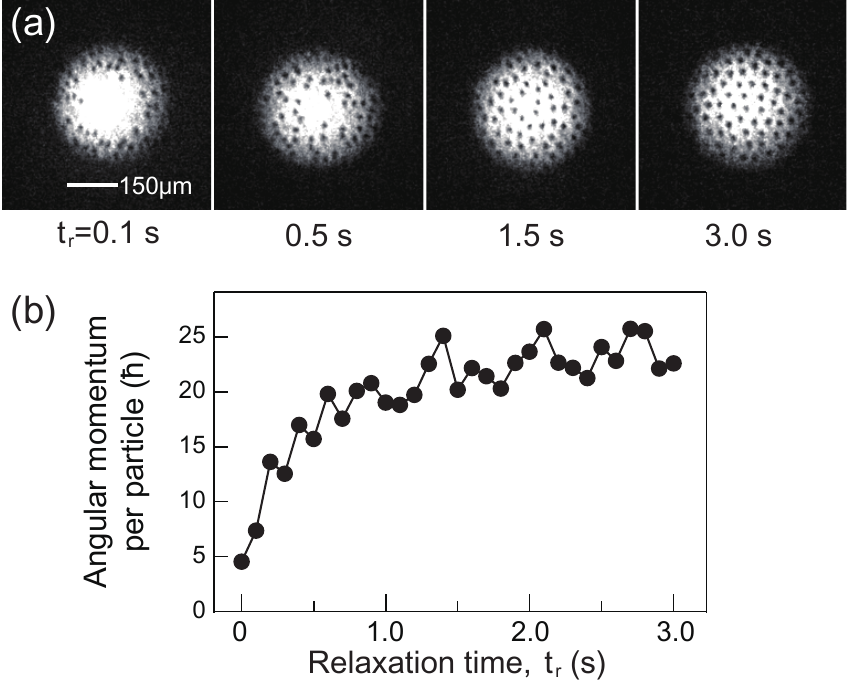}
\caption{Spinning up the condensate by a rotating thermal cloud. (a) Time-of-flight images of the condensate for various hold times $t_r$ after 400~ms driving. There is no noticeable CM motion of the condensate [Fig.~2(b)]. (b) The number of vortices in the condensate gradually increases over the holding time and a vortex lattice forms.}
\end{figure}

The equilibrium process of a condensate in a rotating thermal cloud was theoretically investigated to study the role of a thermal cloud in vortex nucleation and lattice formation~\cite{Williams_prl,Penckwitt02}. Our rotating method provides a new experimental way to prepare a nonequilibrium state where a thermal cloud rotates faster than a condensate. We expect that the rotation frequency difference between the thermal cloud and the condensate may be controlled by the driving parameters or the anharmonicity of the trapping potential. In future, we will pursue  a quantitative analysis of the equilibrium process, including measurements of the rotating frequency of the thermal cloud~\cite{ Haljan_prl,Raman_prl}.

Finally, we present the vortex number $N_v$ after 3~s relaxation as a function of the driving time $t_d$ in Fig.~8. As the driving time increases, the vortex number increases and becomes saturated to $N_v=100$ with damped oscillations. The oscillations reflect the oscillating behavior of the external angular momentum of the driven condensate [Fig.~2(c)]. The total atom number and the thermal fraction of the sample also oscillate accordingly due to the atom loss and heating associated with the relaxation process.

\begin{figure}
\includegraphics[width=7.0cm]{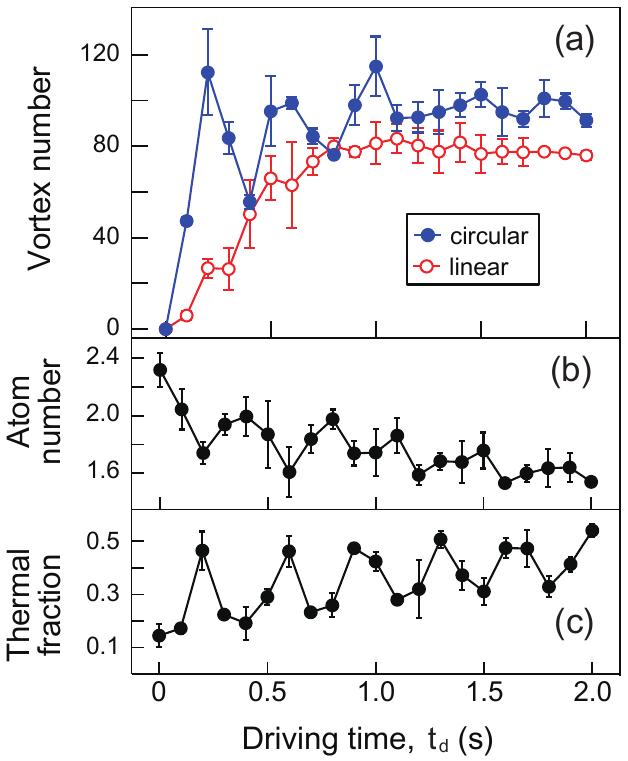}
\caption{(Color online) (a) Vortex number $N_v$ versus driving time $t_d$. The vortex number was measured after 3~s relaxation. The open circles show the result for linear driving, where the transverse bias fields, $B_x$ and $B_y$, were modulated in phase with $\Delta t=0$ [Fig.~2(a)]. Each data point consists of three measurements. The corresponding evolutions of (b) the total atom number and (c) the thermal fraction of the sample.}
\end{figure}

The angular momentum per particle of the stably rotating condensate with 100 vortices is about $40\hbar$, which is only 10\% of the maximum external angular momentum of the circulating condensate. The poor efficiency of the angular momentum transfer to the internal rotation of the condensate is an intrinsic aspect in the evolution of an isolated system. When the energy of the system is given as $E=\frac{1}{2}L_z^2/I$, where $I$ is the moment of inertia of the system, the energy and angular momentum conservations require that $I$ should be preserved in the evolution. This means that when a rotating condensate forms at the trap center, generation of thermal atoms having high angular momentum is unavoidable. This argument suggests that in order to improve the transfer efficiency of the angular momentum, i.e., to obtain a faster rotating condensate, one needs to provide an additional work by dynamically deforming the trapping potential during the relaxation.

\section{Gauge Field Effect}

When a particle with spin moves in a spatially varying magnetic field, the particle acquires a quantum mechanical phase associated with the adiabatic spin rotation, which is known as the Berry phase~\cite{Berry_ProcRoyal}. A gauge potential can be defined for the spatial distribution of the spin orientation and this results in generation of an effective magnetic field for the particle~\cite{GeometricW,Stern}. When a Bose-Einstein condensate is trapped in a magnetic potential, its spin texture is imposed by the magnetic field distribution. Because of the spin-gauge symmetry, the condensate would experience an effective magnetic field originated from the spin texture~\cite{Ho_prl}.

In the current setup, the condensate is radially confined by the quadrupole magnetic field. Its spin texture exhibits a fountain-like configuration which is often referred to as a Skyrmion spin texture and the resultant effective magnetic field is given as
\begin{equation}
\mathbf{B}^e(\mathbf{r}')=\frac{2\hbar}{(r'^2+4z_0^2)^{3/2}}(-\mathbf{r}'+z_0\hat{z}).
\end{equation}
This gauge field is very weak and the corresponding cyclotron frequency, $\omega_c=\hbar/4mz_0^2$, is three orders of magnitude smaller than the trapping frequency~[Fig.~2(c)]. In our recent work~\cite{Choi_prl}, we succeeded in demonstrating the existence of the effective Lorentz force in the current setup by observing that a circular CM motion of the trapped condensate is generated with linear driving, leading to vortex nucleation~[Fig.~8(a)].

In this section, we study the gauge field effect by investigating the response of the condensate to the trap shaking for various elliptical polarizations of the driving. In order to control the polarization of the driving in a fine manner, we reduce $B_y$ by a factor of 6 and apply a rotating bias field as $\{B_x(t),B_y(t)\}=\delta B \{ \sin(\omega_m t), \frac{1}{6} \sin(\omega_m t+\phi)\}$. The relative phase $\phi$ is controlled by changing the turn-on time of $B_y$ as $\Delta t=(3\pi -\phi)/\omega_m$ [Fig.~2(a)].

\begin{figure}
\includegraphics[width=8.0cm]{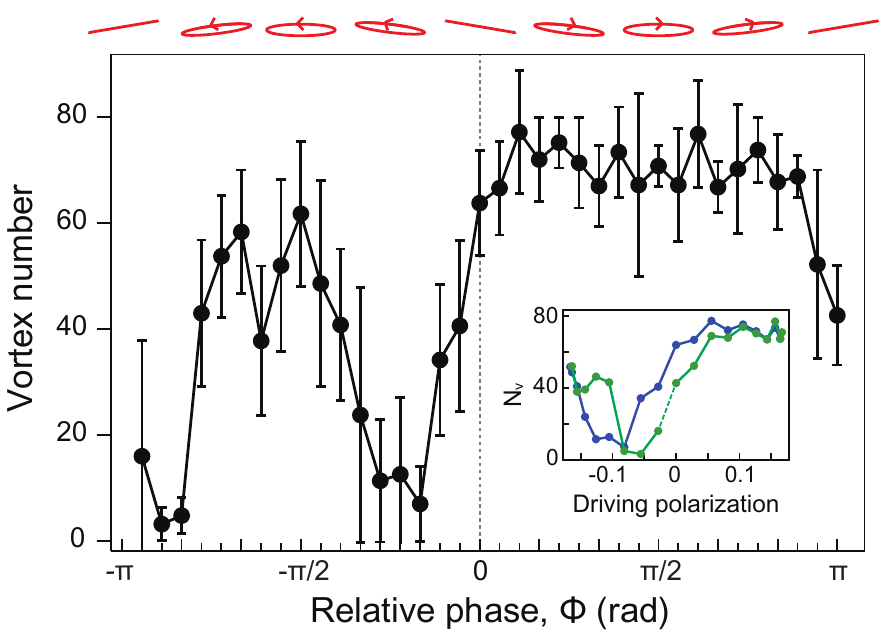}
    \caption{(Color online) Gauge field effect. The vortex number $N_v$ was measured after 500~ms driving and 3~s relaxation for various elliptical polarizations of the driving (see the top axis).  The driving polarization  is controlled by the relative phase $\phi$ between the modulations of $B_x$ and $B_y$ (see text for details). Each data point was obtained from five measurements. The inset displays the same result as a function of the elliptical polarization of the driving. The blue (green) circles denote the data with $|\phi| \leq \pi/2$ ($|\phi|>\pi/2$).}
\end{figure}	

Figure 9 displays the vortex number measured after 500~ms driving and 3~s relaxation as a function of the relative phase. In the result, we see a two-dome structure which is asymmetric with respect to the sign of the relative phase $\phi$. The right dome with clockwise driving is relatively more enhanced than the left dome with counterclockwise driving. This is consistent with the sign of the effective magnetic field. The rotation direction of the condensate was confirmed to be the same as the sign of the driving polarization at $\phi=\pm\pi/2$ by measuring the precession direction of the quadrupole excitations in the rotating condensate~\cite{Zambelli98,Chevy00}.

The vortex generation is almost suppressed at nonzero negative values of $\phi$. An interesting question would be whether we can derive a quantitative relation between the magnitude of the gauge field and the critical driving polarization for nullifying the vortex generation. In the inset of Fig.~9, we replot the data as a function of the driving polarization, $\epsilon_d=\tan (\frac{1}{2} \sin^{-1} (\frac{2 X_m Y_m \sin \phi}{X_m^2+Y_m^2}))$. In our experiment, one driving polarization can be realized with two values of $\phi$ and we see that the vortex numbers for two realizations are slightly different from each other. This means that around the critical driving polarization, the condensate dynamics is sensitive to the turn-on time of $B_y$, i.e., the initial condition of the condensate. A more elaborated setup is desirable for a quantitative study of the gauge field with the critical polarization.

\section{Summary}

We have described a method for rotating a trapped BEC in an anharmonic axisymmetric magnetic potential. By resonantly shaking the trapping potential, we injected a large angular momentum into a system in the form of a circular CM motion of the condensate. We observed that the circulating condensate undergoes rotating shape deformation and eventually relaxes to a rotating condensate having a vortex lattice. We provided a thermal distillation description of the relaxation process. Finally, we demonstrated the gauge field effect by investigating the response of the condensate to elliptical driving with various polarizations.

One of the interesting extensions of this work would be introducing an additional trapping potential during the relaxation to mitigate the centrifugal effect and to increase the rotation speed of the condensate. We expect that our trapping potential might be deformed into a ring potential after the driving phase by applying an additional optical potential~\cite{Ryu07,Ramananth11,Beattie13} or using RF-field dressing techniques~\cite{Lesanovsky07,Sherlock11}, which would result in a giant vortex state~\cite{Lundh_pra,Kavoulakis_NJP}. Recently, Roncaglia \textit{et al.}~\cite{Roncaglia11} showed that a giant vortex state having a large angular momentum in a ring potential is adiabatically connected to the bosonic $\nu=1/2$ Laughlin state in a harmonic potential.

\begin{acknowledgements} 

This work was supported by the NRF of Korea (Grants No.~2011-0017527, No.~2008-0062257, No.~2013-H1A8A1003984).

\end{acknowledgements}


\begin{references}
\bibitem{footnote3} The collision time $\tau=(n_{th} \sigma \bar{v})^{-1}$, where $n_{th}=2.612(m k_B T/2\pi \hbar^2)^{3/2}$ is the atom density of a saturated thermal cloud, $\sigma=8\pi a^2$ denotes the elastic collision cross section (the scattering length $a=2.75$~nm~\cite{Tiesinga96}), and $\bar{v}=4\sqrt{k_BT/\pi m}$ is the thermal velocity.
\bibitem{footnote} In the Thomas-Fermi approximation, we estimated the angular momentum per particle of the condensate as $L_z=\hbar \Sigma_i [1-(r_i/R)^2]^{5/2}$~\cite{BECbook}, where $r_i/R$ is the ratio of the vortex radial position to the condensate radius, which was measured for each vortex from the image.
\end{references}
\end{document}